\documentclass[useAMS,usenatbib]{mn2e}
\usepackage{amssymb}
\usepackage{amsmath}
\usepackage{graphicx}

\title[]{Asteroseismology of the DAV star R808}
\author[Y. H. Chen, H. Shu]{Y. H. Chen$^{1,2,3}$\thanks{E-mail: yanhuichen1987@126.com} and H. Shu$^{1,2}$\\
$^{1}$Institute of Astrophysics, Chuxiong Normal University, Chuxiong 675000, China\\
$^{2}$School of Physics and Electronical Science, Chuxiong Normal University, Chuxiong 675000,China\\
$^{3}$Key Laboratory for the Structure and Evolution of Celestial Objects, Chinese Academy of Sciences, P.O. Box 110, Kunming 650011, China\\}

\begin{document}

\date{Accepted: }

\pagerange{\pageref{firstpage}--\pageref{lastpage}} \pubyear{????}

\maketitle

\label{firstpage}

\begin{abstract}

The DAV star R808 was observed by 13 different telescopes for more than 170 hours in April 2008 on the WET run XCOV26. 25 independent pulsation frequencies were identified by this data set. We assumed 19 $m$ = 0 modes and performed an asteroseismological study on those 19 modes. We evolve grids of DAV star models by \texttt{WDEC} adopting the element diffusion scheme with pure and screened Coulomb potential effect. The core compositions are from white dwarf models evolved by \texttt{MESA}, which are thermal nuclear burning results. Our best fitting model is from the screened Coulomb potential scenario, which has parameters of log($M_{\rm He}/M_{\rm *}$) = -2.4, log($M_{\rm H}/M_{\rm *}$) = -5.2, $T_{\rm eff}$ = 11100\,K, $M_{\rm *}$ = 0.710\,$M_{\odot}$, log$g$ = 8.194, and $\sigma_{RMS}$ = 2.86\,s. The value of $\sigma_{RMS}$ is the smallest among the four existing asteroseismological work. The average period spacing is 46.299\,s for $l$ = 1 modes and 25.647\,s for $l$ = 2 modes. The other 6 observed modes can be fitted by $m$ $\neq$ 0 components of some modes for our best fitting model. Fitting the 25 observed modes, we obtain a $\sigma_{RMS}$ value of 2.59\,s. Considering the period spacings, we also assume, that at least in one case, we detect an $l$ = 2 trapped mode.

\end{abstract}


\section{Introduction}

Pulsating white dwarfs are $g$-mode nonradial pulsators with buoyancy acting as restoring force. There are three main groups of pulsating white dwarfs named as GW Vir stars (PNNV stars and DOV stars), V777 Her stars (DBV stars), and ZZ Ceti stars (DAV stars). The GW Vir stars have stellar atmospheres dominated by He\texttt{II}, C and O lines. The DBV stars have stellar atmospheres dominated by He\texttt{I} lines and the DAV stars have stellar atmospheres dominated by H\texttt{I} lines. The instability strip for GW Vir stars, DBV stars, and DAV stars are around 100\,000 K, 25\,000 K, and 12\,000 K respectively. There are around 20 GW Vir stars, 40 DBV stars, and 250 DAV stars observed (C$\acute{o}$rsico et al. 2019). A large number of DAV stars provide rich research objects for asteroseismology.

Asteroseismology is a powerful tool to detect the inner structure of stars by comparing the observed modes with the calculated ones. For nonradial oscillations, an eigen-mode is indicated by 3 indices, $k$, $l$, and $m$ (Handler 2013). The index $k$ means the order of radial standing wave. The indices $l$ and $m$ are the subscripts of spherical harmonic function, where $l$ is the spherical harmonic degree and $m$ is the azimuthal order. Because of the geometric cancellation effect (Dziembowski 1977), the large $l$ modes have small amplitudes and are not easy to be detected. Therefore, the $l$ values of most observed modes for DAV stars are usually 1 or 2 (Castanheira \& Kepler 2008, and references therein). Due to the rotational frequency splitting, $l$ = 1 modes show triplets and $l$ = 2 modes show quintuplets. Brickhill (1975) derived an approximate relation between the rotation frequency ($\Omega$) and corresponding frequency splitting values ($\delta\nu_{k,l,m}$) for high-overtone ($k$ $>>$ l) modes and slow rotation as
\begin{equation}
\delta \nu_{k,l,m} = \delta m (1-C_{k,l}) \Omega,
\end{equation}
\noindent where the coefficient $C(k,l)$ $\approx$ 1/($l$($l$+1)) for $g$-modes. There is a summary on the regularities between the periods and frequencies of $g$- and $p$-modes in Bogn$\acute{a}$r et al. (2015). In Eq.\,(1), $m$ are integers from -$l$ to $l$, being -1, 0, 1 for $l$ = 1 modes and -2, -1, 0, 1, 2 for $l$ = 2 modes.

On the contrary, the observed triplet modes can be identified as $l$ = 1 modes and the observed quintuplet modes can be identified as $l$ = 2 modes. In addition, the triplet frequency splitting is related to the quintuplet frequency splitting by
\begin{equation}
\frac{\delta\nu_{k,1}}{\delta\nu_{k,2}}=0.6.
\end{equation}
\noindent The relationship is very important for mode identifications. The theory of asteroseismology indicates that the high $k$ g-modes have almost uniform period spacings for successive $k$ modes (same $l$ values). Winget et al. (1981) first proposed that the trapped modes would jump out of the uniform period spacing queue and produce smaller period spacings. These basic laws are helpful for mode identifications. Many pulsating white dwarfs exhibit combination frequencies, which are considered to be derived from the non-linear mixing of sinusoidal signals associated with the mother frequencies. When the stellar convective zone undergos pulsations, the depth of the convective zone will change with the temperature. Therefore, the combination frequencies have the opportunity to be used for mode identifications (Wu 2001, Kerkwijk et al. 2000).

In order to minimize the gaps in the data sets of rapid variable stars caused by the rotation of the Earth, Nather et al. (1990) organized the Whole Earth Telescope (WET). This instrument provides data of continuity for pulsating white dwarfs. The WET runs, together with some other multi-site observations, have increased the independent observed modes of some DAV stars from a few to a dozen or even more, such as G29-38 (Kleinman et al. 1998), EC14012-1446 (Provencal et al. 2012), HL Tau 76 (Dolez et al. 2006), G38-29 and R808 (Thompson et al. 2009), and so on. The rich observed modes are both opportunities and challenges for model fittings. A large number of reliable identified modes can undoubtedly constrain the fitting model more effectively. However, there are many observed modes that are not easy to be identified. In particular, the modes with small differences in frequencies bring challenges to the mode identification, and increase the uncertainty of model fittings.

The DAV star R808, as one of the WET targets, was observed by 13 different telescopes for more than 170 hours between April 4 and April 17, 2008. Thompson et al. (2009) analyzed the time series photometry on R808 and identified 25 independent pulsation modes. In this paper, we present the preliminary asteroseismological study of R808. In Sect. 2, we perform an analysis of the observed modes of R808. We show the input physics and model calculations in Sect. 3. The model fitting work is displaced in Sect. 4 and the analysis of the fitting results is displaced in Sect. 5. Then, we give a discussion and conclusions in Sect. 6.

\section{An Analysis of the Observed Modes of R808}

\begin{table}
\begin{center}
\caption{The observed modes of R808 from Thompson et al. (2009). Fre. is the frequencies in $\mu$Hz and Per. is the corresponding periods in seconds. The frequency intervals of adjacent modes are marked as $\delta$ F in $\mu$Hz. The $P_{sel}$ is the selected periods to constrain the fitting models.}
\begin{tabular}{lllllll}
\hline
ID             &Fre.($\mu$Hz)  &$\delta$ F($\mu$Hz) &Per.(s)          &$P_{sel}$(s)                           \\
\hline
$f_{25}$       &\, 2472.45          &          &\, 404.457            &\, 404.457                             \\
$f_{24}$       &\, 1955.93          &          &\, 511.266            &\, 511.266                             \\
$f_{23}$       &\, 1589.25          &          &\, 629.228            &                                       \\
               &                    &\,7.42    &                      &\, 629.228 or 632.179                  \\
$f_{22}$       &\, 1581.83          &          &\, 632.179            &                                       \\
$f_{21}$       &\, 1342.07          &          &\, 745.120            &\, 745.120                             \\
$f_{20}$       &\, 1255.88          &          &\, 796.253            &\, 796.253                             \\
$f_{19}$       &\, 1186.65          &          &\, 842.707            &\, 842.707                             \\
$f_{18}$       &\, 1162.48          &          &\, 860.227            &\, 860.227                             \\
$f_{17}$       &\, 1142.67          &          &\, 875.146            &                                       \\
               &                    &\,4.34    &                      &\, 875.146 or 878.479                  \\
$f_{16}$       &\, 1138.33          &          &\, 878.479            &                                       \\
$f_{15}$       &\, 1112.71          &          &\, 898.707            &\, 898.707                             \\
$f_{14}$       &\, 1100.81          &          &\, 908.422            &                                       \\
               &                    &\,3.76    &                      &\, 908.422 or 911.534                  \\
$f_{13}$       &\, 1097.05          &          &\, 911.534            &                                       \\
               &                    &\,3.78    &                      &                                       \\
$f_{12}$       &\, 1093.27          &          &\, 914.683            &                                       \\
               &                    &\,1.33    &                      &                                       \\
$f_{11}$       &\, 1091.94          &          &\, 915.803            &\, 914.683 or 915.803                  \\
               &                    &\,7.93    &                      &\, or/and 922.504                      \\
$f_{10}$       &\, 1084.01          &          &\, 922.504            &                                       \\
$f_{09}$       &\, 1049.98          &          &\, 952.398            &                                       \\
               &                    &\,8.88    &                      &\, 952.398                             \\
               &                    &          &                      &\, or/and 960.527                      \\
$f_{08}$       &\, 1041.10          &          &\, 960.527            &                                       \\
$f_{07}$       &\,  988.736         &          &\,1011.39             &\,1011.39                              \\
$f_{06}$       &\,  961.476         &          &\,1040.07             &                                       \\
               &                    &\,1.827   &                      &\,1040.07 or 1042.05                   \\
$f_{05}$       &\,  959.649         &          &\,1042.05             &                                       \\
$f_{04}$       &\,  937.444         &          &\,1066.73             &\,1066.73                              \\
$f_{03}$       &\,  916.516         &          &\,1091.09             &\,1091.09                              \\
$f_{02}$       &\,  874.160         &          &\,1143.96             &\,1143.96                              \\
$f_{01}$       &\,  406.652         &          &\,2459.10             &\,2459.10                              \\
\hline
\end{tabular}
\end{center}
\end{table}

The DA white dwarf R808 was first identified by McGraw et al. as a DAV star through four nights observations in 1975 at McDonald Observatory (McGraw et al. 1976). R808 was one of the targets of the WET run denoted as XCOV26 in April 2008 (Thompson et al. 2009). The WET runs, together with multi-site observation campaigns, planned to measure the light curve and the corresponding eigen frequencies accurately. The light curve and the shapes of pulses are used to study the relationship between the depth of the convention zone and the temperature. The eigen frequencies are used to the study of asteroseismology.

The theoretical studies show that the depth of the convective zone varies with the temperature of the star and strongly influences the modes. Therefore, pronounced nonlinearities and combination modes are present. The combination modes are helpful to the mode identifications. However, there were only 3 combination modes identified by Thompson et al. (2009) for R808. The convection zone fitting technique needs the asteroseismological mode identifications on $l$ and $m$ values.

Thompson et al. (2009) identified 25 independent frequencies, which are list in Table 1 in this paper. We calculate the frequency intervals of some adjacent modes, which are marked as $\delta$ F in $\mu$Hz in Table 1. The $\delta$ F values of 3.76\,$\mu$Hz and 3.78\,$\mu$Hz are very close. But there is no $\delta$ F values of 3.77/0.6 = 6.28$\mu$Hz or 3.77*0.6 = 2.26$\mu$Hz. At last, we assume most of the observed modes are $m$ = 0 , $l$ = 1 or 2 modes. The fifth column is the selected periods to constrain the fitting models. For periods close in frequencies, we take any one to constrain the fitting models, such as $f_{23}$ or $f_{22}$. However, for two periods which are not very close in frequencies, we take one or two to constrain the fitting models, such as $f_{09}$ or/and $f_{08}$. The modes in the fifth column of Table 1 are used to constrain the fitting models.

\section{Input Physics and Model Calculations}

\begin{table}
\begin{center}
\caption{The grid size and steps table.}
\begin{tabular}{lllll}
\hline
Parameters                  &grid                  &pre                 &refined                             \\
                            &-size                 &-steps              &-steps                              \\
\hline
$M_{*}$/$M_{\odot}$         &0.56to0.72            &0.01                &0.005                               \\
$T_{\rm eff}$(K)            &10800to11800          &200                 &50                                  \\
log($M_{\rm He}/M_{\rm *}$) &-2.0to-4.0            &0.5                 &0.1                                 \\
log($M_{\rm H}/M_{\rm *}$)  &-4.0to-10.0           &1.0                 &0.1                                 \\
\hline
\end{tabular}
\end{center}
\end{table}

The fifth column of Table 1 contains abundant observed modes. We plan to fit these observed modes using two groups of calculated modes, namely two groups of DAV star models. One group of DAV star models are evolved using the pure Coulomb potential. The other group of DAV star models are evolved using the screened Coulomb potential.

In this work, we evolve a group of main-sequence stars to be hot white dwarfs by \texttt{MESA} code, version number 4298 (Paxton et al. 2011). These white dwarf models have thermal nuclear burning core compositions. We take the core compositions and add them to the seed models of an older version of \texttt{WDEC}. \texttt{WDEC} was first developed by Schwarzschild, which was designed to calculate the evolutions of white dwarfs without previous stellar evolutions (Kutter \& Savedoff 1969, Lamb \& van Horn 1975, and Wood 1990). \texttt{WDEC} does not calculate the nuclear reaction and mass-loss processes. Therefore, thermal nuclear burning core compositions are helpful to the seed models of \texttt{WDEC}. For \texttt{WDEC}, the equation of state (EOS) are derived by Lamb (1974) and Saumon et al. (1995). The opacities are performed by Itoh et al. (1983, 1984). There is a new version of \texttt{WDEC} (Bischoff-Kim \& Montgomery 2018), which directly adopts EOS and opacities of \texttt{MESA} (version number 8118) and contains the stellar oscillation routines. We will apply the new version of \texttt{WDEC} to do asteroseismological research on white dwarfs in the future.

\texttt{WDEC} treats the composition transition zones (C/He and He/H) as the diffusion equilibrium profiles. Su et al. (2014) added the element diffusion scheme of Thoul, Bahcall \& Loeb (1994) into \texttt{WDEC} to evolve the time dependent element diffusion DAV star models. The Coulomb logarithm equation is for a pure Coulomb potential with a cutoff at the Debye radius (Thoul, Bahcall \& Loeb 1994). Therefore, we mark the method to evolve DAV star models as the pure Coulomb potential scenario.

Thoul, Bahcall \& Loeb (1994) used the element diffusion scheme to study the solar interior. Paquette et al. (1986) considered that a screened Coulomb potential would be more suitable for white dwarfs. The Burgers equations of momentum and energy conservation, and the equation of Coulomb logarithm are revised, according to the equations (1, 2, 22-25) of Muchmore (1984). The method adopts the screened Coulomb potential to calculate the element diffusion processes (Cox, Guzik \& Kidman 1989). We mark the method to evolve DAV star models as the screened Coulomb potential scenario. For more details, see Chen (2018, 2020).

Two groups of DAV star models are evolved with the methods above. \texttt{WDEC} calculates the C/He and He/H transition zones with pure or screened Coulomb potential. The mixing length parameter is adopted to be 0.6 (Bergeron et al. 1995). Each group of DAV star models are four-parameter grid models, total stellar mass in solar mass $M_{*}$/$M_{\odot}$, effective temperature $T_{\rm eff}$, logarithm of helium mass fraction log($M_{\rm He}/M_{\rm *}$), and logarithm of hydrogen mass fraction log($M_{\rm H}/M_{\rm *}$). The grid size and steps are shown in Table 2. We calculate the theoretical pulsation modes based on the pulsation code of Li (1992a, b).

An equation of root-mean-square residual $\sigma_{RMS}$ is used to evaluate the fitting results.
\begin{equation}
\sigma_{RMS}=\sqrt{\frac{1}{n} \sum_{n}(P_{\rm obs}-P_{\rm cal})^2}.
\end{equation}
\noindent In Eq.\,(3), $n$ denotes the number of fitted observed modes. We reduce the parameter steps around the models with minimum $\sigma_{RMS}$ values, as shown in the fourth column of Table 2. At last, a best-fitting model will be selected.

\section{The Model Fitting Work of R808}

\begin{table*}
\begin{center}
\caption{The best fitting model parameters for fitting 16 combinations of a total of 19 observed modes.}
\begin{tabular}{lllllllllllllll}
\hline
Combinations of observed modes      &$T_{\rm eff}$(K)&$M_{*}$/$M_{\odot}$  &log($M_{\rm H}/M_{\rm *}$)     &log($M_{\rm He}/M_{\rm *}$)&$\sigma_{RMS}$(s)\\
\hline
15 observed modes added             &(pure/screened) &(pure/screened)      &(pure/screened)                &(pure/screened)            &(pure/screened)  \\
\hline
$f_{23}$, $f_{17}$, $f_{14}$, $f_{06}$ &11150/11150  &0.705/0.700          &-5.2/-5.0                      &-2.3/-2.4                  &3.11/2.96        \\
$f_{22}$, $f_{17}$, $f_{14}$, $f_{06}$ &11150/11150  &0.705/0.700          &-5.2/-5.0                      &-2.3/-2.4                  &3.24/3.03        \\
$f_{23}$, $f_{16}$, $f_{14}$, $f_{06}$ &11150/11100  &0.705/0.710          &-5.2/-5.2                      &-2.3/-2.4                  &3.10/2.96        \\
$f_{22}$, $f_{16}$, $f_{14}$, $f_{06}$ &11150/11250  &0.705/0.700          &-5.2/-5.1                      &-2.3/-2.3                  &3.23/3.07        \\
$f_{23}$, $f_{17}$, $f_{13}$, $f_{06}$ &11150/11100  &0.705/0.710          &-5.2/-5.2                      &-2.3/-2.4                  &3.09/2.87        \\
$f_{22}$, $f_{17}$, $f_{13}$, $f_{06}$ &11150/11100  &0.705/0.710          &-5.2/-5.2                      &-2.3/-2.4                  &3.22/2.98        \\
$f_{23}$, $f_{16}$, $f_{13}$, $f_{06}$ &11150/11100  &0.705/0.710          &-5.2/-5.2                      &-2.3/-2.4                  &3.08/2.86        \\
$f_{22}$, $f_{16}$, $f_{13}$, $f_{06}$ &11150/11100  &0.705/0.710          &-5.2/-5.2                      &-2.3/-2.4                  &3.21/2.98        \\
$f_{23}$, $f_{17}$, $f_{14}$, $f_{05}$ &11150/11150  &0.705/0.700          &-5.2/-5.0                      &-2.3/-2.4                  &3.24/3.06        \\
$f_{22}$, $f_{17}$, $f_{14}$, $f_{05}$ &11150/10950  &0.705/0.705          &-5.2/-5.0                      &-2.3/-2.5                  &3.37/3.08        \\
$f_{23}$, $f_{16}$, $f_{14}$, $f_{05}$ &11150/11100  &0.705/0.705          &-5.2/-5.2                      &-2.3/-2.3                  &3.23/3.07        \\
$f_{22}$, $f_{16}$, $f_{14}$, $f_{05}$ &11150/10850  &0.705/0.700          &-5.2/-4.8                      &-2.3/-2.5                  &3.36/3.09        \\
$f_{23}$, $f_{17}$, $f_{13}$, $f_{05}$ &11150/11100  &0.705/0.710          &-5.2/-5.2                      &-2.3/-2.4                  &3.22/2.99        \\
$f_{22}$, $f_{17}$, $f_{13}$, $f_{05}$ &11150/11100  &0.705/0.710          &-5.2/-5.2                      &-2.3/-2.4                  &3.35/3.10        \\
$f_{23}$, $f_{16}$, $f_{13}$, $f_{05}$ &11150/11100  &0.705/0.710          &-5.2/-5.2                      &-2.3/-2.4                  &3.21/2.99        \\
$f_{22}$, $f_{16}$, $f_{13}$, $f_{05}$ &11100/11100  &0.710/0.710          &-5.2/-5.2                      &-2.4/-2.4                  &3.33/3.10        \\
\hline
\end{tabular}
\end{center}
\end{table*}

\begin{figure}
\begin{center}
\includegraphics[width=9.0cm,angle=0]{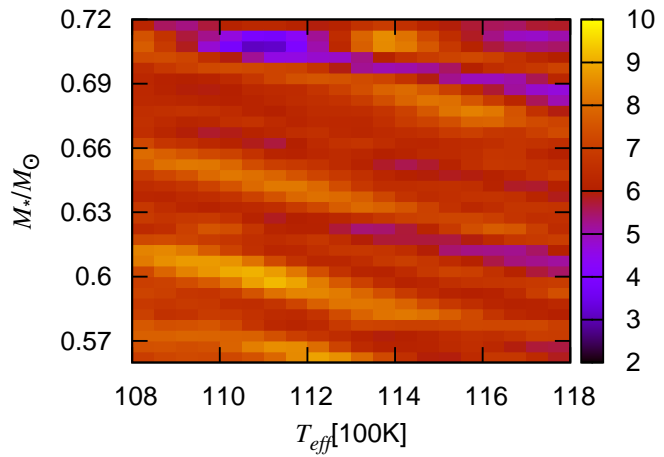}
\end{center}
\caption{The color residual diagram to fit R808 with DAV star models. Those models are from the screened Coulomb potential scenario with log($M_{\rm He}/M_{\rm *}$) = -2.4 and log($M_{\rm H}/M_{\rm *}$) = -5.2. The colors denote the values of $\sigma_{RMS}$.}
\end{figure}

\begin{table}
\begin{center}
\caption{The table of detailed fitting results. The model parameters are log($M_{\rm He}/M_{\rm *}$) = -2.4, log($M_{\rm H}/M_{\rm *}$) = -5.2, $T_{\rm eff}$ = 11100\,K, $M_{\rm *}$ = 0.710\,$M_{\odot}$, and log$g$ = 8.194. $P_{obs}$ is the observed modes, $P_{cal(p)}$/$P_{cal(s)}$ is the calculated modes through the pure/screened Coulomb potential scenario.}
\begin{tabular}{lllll}
\hline
$P_{obs}$    &$P_{cal(p)}$             &$P_{obs}$-$P_{cal(p)}$        &$P_{cal(s)}(l,k)$    &$P_{obs}$-$P_{cal(s)}$              \\
\hline
(s)          &(s)                      &(s)                           &(s)                  &(s)                                 \\
\hline
404.457      &\, 400.138               &$\,$ 4.319                    &\,400.261(2,12)      &$\,$ 4.196                          \\
511.266      &\, 518.710               &$\,$-7.444                    &\,517.447(1, 9)      &$\,$-6.181                          \\
629.228      &\, 628.784               &$\,$ 0.444                    &\,628.516(1,11)      &$\,$ 0.712                          \\
745.120      &\, 746.605               &$\,$-1.485                    &\,744.469(2,26)      &$\,$ 0.651                          \\
796.253      &\, 794.457               &$\,$ 1.796                    &\,793.080(2,28)      &$\,$ 3.173                          \\
842.707      &\, 848.854               &$\,$-6.147                    &\,847.160(2,30)      &$\,$-4.453                          \\
860.227      &\, 860.062               &$\,$ 0.165                    &\,858.589(1,17)      &$\,$ 1.638                          \\
878.479      &\, 877.558               &$\,$ 0.921                    &\,876.888(2,31)      &$\,$ 1.591                          \\
898.707      &\, 900.077               &$\,$-1.370                    &\,898.874(2,32)      &$\,$-0.167                          \\
911.534      &\, 911.891               &$\,$-0.357                    &\,911.677(1,18)      &$\,$-0.143                          \\
922.504      &\, 923.219               &$\,$-0.715                    &\,921.378(2,33)      &$\,$ 1.126                          \\
952.398      &\, 952.117               &$\,$ 0.281                    &\,951.120(2,34)      &$\,$ 1.278                          \\
960.527      &\, 967.259               &$\,$-6.732                    &\,964.172(1,19)      &$\,$-3.645                          \\
1011.39      &\,1011.53                &$\,$-0.14                     &\,1010.66(2,36)      &$\,$ 0.73                           \\
1040.07      &\,1039.07                &$\,$ 1.00                     &\,1037.53(1,21)      &$\,$ 2.54                           \\
1066.73      &\,1063.31                &$\,$ 3.42                     &\,1061.63(2,38)      &$\,$ 5.10                           \\
1091.09      &\,1093.54                &$\,$-2.45                     &\,1092.17(2,39)      &$\,$-1.08                           \\
             &\,1096.04                &$\,$-4.95                     &\,1093.96(1,22)      &$\,$-2.87                           \\
1143.96      &\,1142.11                &$\,$ 1.85                     &\,1140.92(2,41)      &$\,$ 3.04                           \\
2459.10      &\,2460.75                &$\,$-1.65                     &\,2457.04(1,51)      &$\,$ 2.06                           \\
\hline
$\sigma_{RMS}$&3.17\,s                 &                              &2.86\,s              &                                    \\
\hline
\end{tabular}
\end{center}
\end{table}

\begin{figure}
\begin{center}
\includegraphics[width=9.0cm,angle=0]{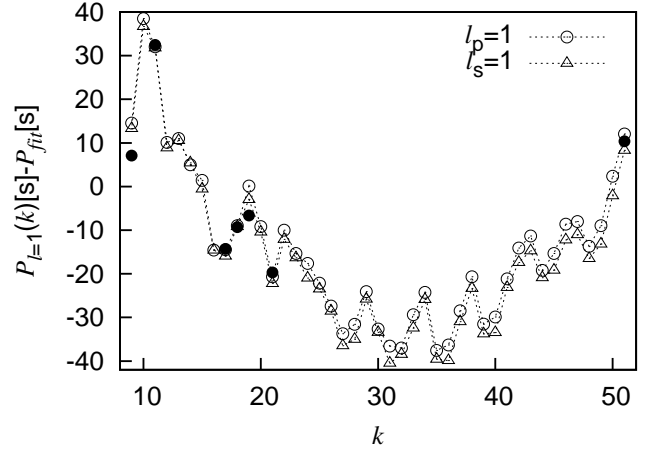}
\end{center}
\caption{The diagram to fit $l$ = 1 modes. The observed and calculated modes are from Table 4. The filled dots denote the observed modes, which are fitted by $P_{fit}$ = 46.2990*$k$ + 87.4714. The open dots are calculated from the pure Coulomb potential scenario, while the triangles are calculated from the screened Coulomb potential scenario. The radial order $k$ is adopted to be from 9 to 51.}
\end{figure}

\begin{figure}
\begin{center}
\includegraphics[width=9.0cm,angle=0]{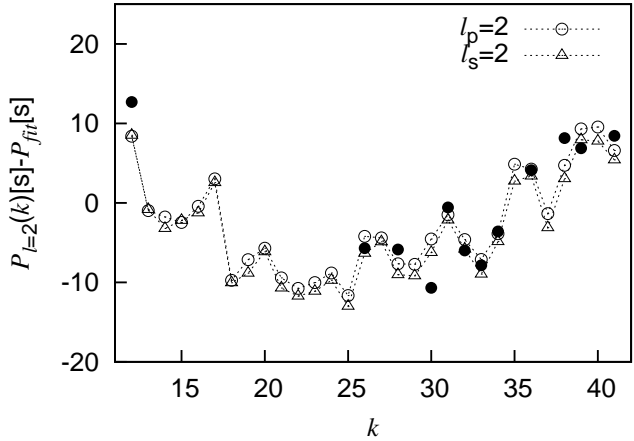}
\end{center}
\caption{The same as Fig. 2, but for $l$ = 2 modes. The fitting function is $P_{fit}$ = 25.6468*$k$ + 84.0009. The radial order $k$ is adopted to be from 12 to 41.}
\end{figure}

\begin{figure}
\begin{center}
\includegraphics[width=9.0cm,angle=0]{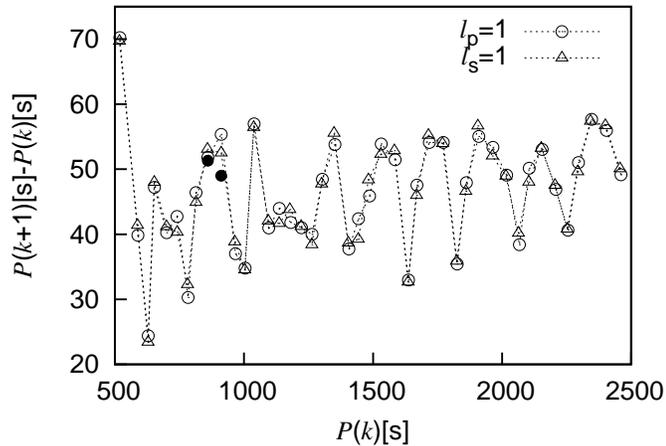}
\end{center}
\caption{The period to period spacing diagram for $l$ = 1 modes. The radial order $k$ is adopted to be from 9 to 51.}
\end{figure}

\begin{figure}
\begin{center}
\includegraphics[width=9.0cm,angle=0]{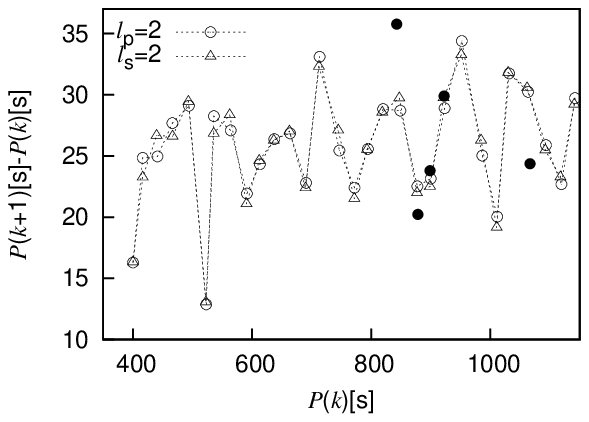}
\end{center}
\caption{The same as Fig. 4, but for $l$ = 2 modes. The radial order $k$ is adopted to be from 12 to 41.}
\end{figure}

The 12 modes of $f_{25}$, $f_{24}$, $f_{21}$ to $f_{18}$, $f_{15}$, $f_{07}$, $f_{04}$ to $f_{01}$ are assumed to be $m$ = 0 modes. First, we fitted the 12 observed modes plus $f_{23}$, $f_{17}$, $f_{14}$, $f_{12}$, $f_{10}$, $f_{09}$, $f_{08}$, and $f_{06}$. The observed modes of $f_{09}$ and $f_{08}$ can be fitted by two theoretical modes. We assume both the observed modes of $f_{09}$ and $f_{08}$ to be $m$ = 0 modes. However, the observed modes of $f_{14}$ and $f_{12}$ are fitted by the same theoretical mode, which is obviously closer to the observed mode of $f_{14}$. We assume both the observed modes of $f_{12}$ and $f_{11}$ to be $m$ $\neq$ components and the observed mode of $f_{10}$ to be $m$ = 0 mode. Then, we have the 12 observed modes plus $f_{09}$, $f_{08}$, and $f_{10}$, a total of 15 observed modes with $m$ = 0. The 15 observed modes plus $f_{23}$ or $f_{22}$, $f_{17}$ or $f_{16}$, $f_{14}$ or $f_{13}$, $f_{06}$ or $f_{05}$, a total of 19 observed modes were used to constrain the fitting models. Therefore, $n$ is considered to be 19 in Eq.\,(3).

We use the 15 observed modes plus $f_{23}$, $f_{17}$, $f_{14}$, and $f_{06}$ to preliminarily constrain the fitting models. The pre-best fitting model is from the screened Coulomb potential scenario, and the parameters of the pre-best fitting model are log($M_{\rm He}/M_{\rm *}$) = -2.5, log($M_{\rm H}/M_{\rm *}$) = -5.0, $T_{\rm eff}$ = 11200\,K, $M_{\rm *}$ = 0.700\,$M_{\odot}$, and $\sigma_{RMS}$ = 3.46\,s. We reduce the parameter steps around the pre-best model. There are 16 combinations of 15 plus 4 observed modes, as shown in the first column of Table 3. Those best fitting model parameters are very close in Table 3. We chose one with the smallest $\sigma_{RMS}$ (2.86\,s) as the final best fitting model, which is from the screened Coulomb potential scenario. The best fitting model parameters are log($M_{\rm He}/M_{\rm *}$) = -2.4, log($M_{\rm H}/M_{\rm *}$) = -5.2, $T_{\rm eff}$ = 11100\,K, $M_{\rm *}$ = 0.710\,$M_{\odot}$, and log$g$ = 8.194. With the fixed helium and hydrogen mass fraction of log($M_{\rm He}/M_{\rm *}$) = -2.4 and log($M_{\rm H}/M_{\rm *}$) = -5.2, we show the color residual diagram to fit R808 in Fig. 1. It is obvious that the best fitting model has $T_{\rm eff}$ = 11100\,K and $M_{\rm *}$ = 0.710\,$M_{\odot}$.

We show the detailed fitting results in Table 4. The observed modes of $f_{23}$, $f_{16}$, $f_{13}$, and $f_{06}$ are assumed to be $m$ = 0 modes, according to the best fitting model. There are 19 observed modes in the first column of Table 4. The second column are the calculated modes through the pure Coulomb potential scenario. The fourth column are the calculated modes through the screened Coulomb potential scenario. We show the spherical harmonic degrees and the radial orders in the brackets. There are 7 $l$ = 1 modes and 12 $l$ = 2 modes. In addition, the $f_{03}$ mode of 1091.09\,s can be fitted by 1092.17\,s ($l$=2, $k$=39) or 1093.96\,s ($l$=1, $k$=22). The corresponding $P_{obs}$ minus $P_{cal}$ values are calculated for each observed mode.

Fitting the 19 observed modes, $\sigma_{RMS}$ is 3.17\,s/2.86\,s for the theoretical modes calculated through the pure/screened Coulomb potential scenario. From the pure Coulomb potential scenario to the screened Coulomb potential scenario, some observed modes are fitted better obviously such as $f_{24}$( 511.266\,s), $f_{19}$ (842.707\,s), $f_{08}$ (960.527\,s) in Table 4, $\sigma_{RMS}$ is improved by 9.78\%.

In Fig. 2, we show the detailed fitting results for $l$ = 1 modes. The radial order $k$ is adopted to be from 9 to 51 in Fig. 2. The average period spacing for $l$ = 1 modes is 46.299\,s. The filled dots are the observed modes, the open dots are the calculated modes from the pure Coulomb potential scenario, the triangles are the calculated modes from the screened Coulomb potential scenario. Figure 3 is the detailed fitting results for $l$ = 2 modes. The radial order $k$ is adopted to be from 12 to 41. The average period spacing for $l$ = 2 modes is 25.647\,s. The ratio between 46.299\,s and 25.647\,s is 1.805, which is close to the theoretical value of $\sqrt{3}$ = 1.732. In Fig. 4 and 5, we show the period to period spacing diagram for $l$ = 1 and 2 modes respectively. For $l$ = 1 modes, $k$ is adopted to be from 9 to 51. For $l$ = 2 modes, $k$ is adopted to be from 12 to 41. In Fig. 4, we can see that the period spacings are around $\sim$45\,s. In Fig. 5, we can see that the period spacings are around $\sim$25\,s.

\section{The Preliminary Analysis of the Fitting Results}

\begin{table*}
\begin{center}
\caption{Table of best fitting models. The ID number 1 is from the spectral results of Bergeron et al. (2004). The ID number 2, 3, 4, and 5 is from the asteroseismological results of Castanheira \& Kepler (2009), Bischoff-Kim (2009), Romero et al. (2012), and the present paper respectively.}
\begin{tabular}{llllllllllllllll}
\hline
ID     &$T_{\rm eff}$  &log\,$g$       &$M_{*}$                &log($M_{\rm H}/M_{*}$)  &log($M_{\rm He}/M_{*}$)&$\sigma_{RMS}$\\
       &(K)            &               &($M_{\odot}$)          &                        &                       &(s)           \\
\hline
1      &11160$\pm$200  &8.04$\pm$0.05  &0.626$\pm$0.028        &                        &                       &              \\
2      &11000          &               &0.65                   &-9.5                    &-3.5                   &5.48          \\
3      &11250          &               &0.675                  &-4.62                   &-2.58                  &3.61          \\
4      &11213$\pm$130  &8.18$\pm$0.05  &0.705$\pm$0.033        &-4.28 to -4.72          &-2.12                  &4.00          \\
5(s)   &11100          &8.194          &0.710                  &-5.2                    &-2.4                   &2.86          \\
\hline
\end{tabular}
\end{center}
\end{table*}

\begin{table*}
\begin{center}
\caption{The detailed fitting results and a possible mode identification for 25 observed modes based on the best fitting model. The value of $\sigma_{RMS}$ is 2.59\,s when fitting the 25 observed modes.}
\begin{tabular}{lllllllllllll}
\hline
$P_{\rm cal(s)}(l,k,m)$&$P_{\rm obs}$&$P_{obs}$-$P_{cal(s)}$&$P_{\rm cal(s)}(l,k,m)$&$P_{\rm obs}$&$P_{obs}$-$P_{cal(s)}$&$P_{\rm cal(s)}(l,k,m)$&$P_{\rm obs}$&$P_{obs}$-$P_{cal(s)}$\\
(s)               &(s)      &(s)         &(s)              &(s)      &(s)         &(s)              &(s)      &(s)         \\
\hline
138.729(1,1,0)    &         &            &160.761(2,3,0)   &         &            &818.592(2,29,0)  &         &            \\
207.873(1,2,0)    &         &            &194.745(2,4,0)   &         &            &847.160(2,30,0)  &842.707  &$\,$-4.453  \\
278.302(1,3,0)    &         &            &207.271(2,5,0)   &         &            &867.333(2,31,+2) &         &            \\
302.893(1,4,0)    &         &            &227.620(2,6,0)   &         &            &872.083(2,31,+1) &875.146  &$\,$ 3.063  \\
340.082(1,5,0)    &         &            &267.739(2,7,0)   &         &            &876.888(2,31,0)  &878.479  &$\,$ 1.591  \\
393.802(1,6,0)    &         &            &289.890(2,8,0)   &         &            &881.741(2,31,-1) &         &            \\
459.064(1,7,0)    &         &            &301.182(2,9,0)   &         &            &886.651(2,31,-2) &         &            \\
479.819(1,8,0)    &         &            &339.863(2,10,0)  &         &            &888.841(2,32,+2) &         &            \\
517.447(1,9,0)    &511.266  &$\,$-6.181  &365.676(2,11,0)  &         &            &893.831(2,32,+1) &         &            \\
587.142(1,10,0)   &         &            &400.261(2,12,0)  &404.457  &$\,$ 4.196  &898.874(2,32,0)  &898.707  &$\,$-0.167  \\
628.516(1,11,0)   &629.228  &$\,$ 0.712  &416.584(2,13,0)  &         &            &903.979(2,32,-1) &         &            \\
651.918(1,12,0)   &         &            &439.837(2,14,0)  &         &            &909.140(2,32,-2) &         &            \\
699.910(1,13,0)   &         &            &466.495(2,15,0)  &         &            &910.838(2,33,+2) &         &            \\
741.080(1,14,0)   &         &            &493.100(2,16,0)  &         &            &916.078(2,33,+1) &915.803  &$\,$-0.275  \\
781.394(1,15,0)   &         &            &522.546(2,17,0)  &         &            &921.377(2,33,0)  &922.504  &$\,$ 1.126  \\
813.677(1,16,0)   &         &            &535.604(2,18,0)  &         &            &926.741(2,33,-1) &         &            \\
858.589(1,17,0)   &860.227  &$\,$ 1.638  &562.435(2,19,0)  &         &            &932.166(2,33,-2) &         &            \\
908.554(1,18,+1)  &908.422  &$\,$-0.132  &590.787(2,20,0)  &         &            &951.120(2,34,0)  &952.398  &$\,$ 1.278  \\
911.677(1,18,0)   &911.534  &$\,$-0.143  &611.878(2,21,0)  &         &            &984.379(2,35,0)  &         &            \\
914.821(1,18,-1)  &914.683  &$\,$-0.138  &631.453(2,22,+2) &632.179  &$\,$ 0.726  &1010.66(2,36,0)  &1011.39  &$\,$ 0.73   \\
964.172(1,19,0)   &960.527  &$\,$-3.645  &633.967(2,22,+1) &         &            &1016.66(2,37,+2) &         &            \\
1003.01(1,20,0)   &         &            &636.500(2,22,0)  &         &            &1023.20(2,37,+1) &         &            \\
1037.53(1,21,0)   &1040.07  &$\,$ 2.54   &639.055(2,22,-1) &         &            &1029.81(2,37,0)  &         &            \\
1093.96(1,22,0)   &         &            &641.630(2,22,-2) &         &            &1036.52(2,37,-1) &         &            \\
1136.01(1,23,0)   &         &            &662.759(2,23,0)  &         &            &1043.31(2,37,-2) &1042.05  &$\,$-1.26   \\
1177.67(1,24,0)   &         &            &689.779(2,24,0)  &         &            &1061.63(2,38,0)  &1066.73  &$\,$ 5.10   \\
2457.04(1,51,0)   &2459.10  &$\,$ 2.06   &712.158(2,25,0)  &         &            &1092.17(2,39,0)  &1091.09  &$\,$-1.08   \\
                  &         &            &744.469(2,26,0)  &745.120  &$\,$ 0.651  &1117.65(2,40,0)  &         &            \\
121.199(2,2,0)    &         &            &771.570(2,27,0)  &         &            &1140.92(2,41,0)  &1143.96  &$\,$ 3.04   \\
121.199(2,2,0)    &         &            &793.080(2,28,0)  &796.253  &$\,$ 3.173  &1170.13(2,42,0)  &         &            \\
\hline
\end{tabular}
\end{center}
\end{table*}

Fitting the 19 observed modes of R808, $\sigma_{RMS}$ is improved by 9.78\% from the pure Coulomb potential scenario to the screened Coulomb potential scenario. Fitting the 6 observed modes of HS 0507+0434B, $\sigma_{RMS}$ was improved by 34\% from the pure Coulomb potential scenario to the screened Coulomb potential scenario (Chen 2020). In Table 4, Fig. 2, and Fig. 3, we can see that most of the observed modes are basically reduced by around 1 second considering the screened Coulomb potential. If there are a large number of calculated modes which have slightly longer periods than the observed modes, the screened Coulomb potential must be helpful.

In Table 5, we show the best fitting models of the spectral results of Bergeron et al. (2004) and the asteroseismological results of Castanheira \& Kepler (2009), Bischoff-Kim (2009), Romero et al. (2012), and the present paper respectively. Castanheira \& Kepler (2009) evolved grids of DAV star models by \texttt{WDEC} with a fixed homogeneous C/O 50:50 core. Bischoff-Kim (2009) calculated the asteroseismological models by \texttt{WDEC} with a parameterized central O abundance and O profiles. Romero et al. (2012) calculated the asteroseismological models by \texttt{LPCODE} with time dependent element diffusion effect. The effective temperature values in Table 5 are basically consistent. The total stellar mass and the gravitational acceleration obtained by this work are basically consistent with those obtained by Romero et al. (2012). The log($M_{\rm H}/M_{*}$) value obtained by this work is slightly smaller than that obtained by Bischoff-Kim (2009) and Romero et al. (2012). The log($M_{\rm He}/M_{*}$) value obtained by this work is in the middle between that obtained by Bischoff-Kim (2009) and Romero et al. (2012). Castanheira \& Kepler (2009) fitted 8 observed modes of R808 and obtained a $\sigma_{RMS}$ value of 5.48\,s. Bischoff-Kim (2009) fitted 18 modes of R808 and obtained a $\sigma_{RMS}$ value of 3.61\,s. Romero et al. (2012) fitted 17 modes of R808 and obtained a $\sigma_{RMS}$ value of 4.00\,s. We fit 19 modes of R808 and obtain a $\sigma_{RMS}$ value of 2.86\,s. The mode identifications (values of $k$, $l$, and $m$) of the observed modes are not exactly same among the four asteroseismological work. Even in this work, the mode of $f_{03}$ (1091.09) can be fitted by 1092.17($l$=2, $k$=39) or 1093.96($l$=1, $k$=22). More efforts are needed for the detailed mode identifications.

We show the detailed fitting results for the best fitting model in Table 6. The modes of $f_{18}$ (860.227\,s), $f_{13}$ (911.534\,s), $f_{08}$ (960.527\,s) are fitted by modes of $l$ = 1, $k$ = 17, 18, and 19 respectively. The corresponding period spacing is 51.307\,s and 48.993\,s respectively, as shown in Fig. 4, which are larger than the average period spacing for $l$ = 1 modes (46.299\,s). The modes of $f_{19}$ (842.707\,s), $f_{16}$ (878.479\,s), $f_{15}$ (898.707\,s), $f_{10}$ (922.504\,s), $f_{09}$ (952.398\,s), $f_{04}$ (1066.73\,s), $f_{03}$ (1091.09\,s) are fitted by modes of $l$ = 2, $k$ = 30, 31, 32, 33, 34, 38, and 39 respectively. The corresponding period spacing is 35.772\,s, 20.228\,s, 23.797\,s, 29.894\,s, and 24.36\,s respectively, as shown in Fig. 5. The average period spacing for the observed $l$ = 2 modes is 25.647\,s. Therefore, the modes of $f_{16}$ (878.479\,s) and/or $f_{15}$ (898.707\,s) may be the trapped modes. However, the thinner the H atmosphere mass, the more obvious the mode trapping effect (Brassard et al. 1992). The H atmosphere mass is not very thin for the last three best fitting models in Table 5 to fit R808.

In Table 4, the mode of $f_{13}$ (911.534\,s) is fitted by an $l$ = 1 mode. Therefore, we assume that $f_{14}$, $f_{13}$, and $f_{12}$ in Table 1 make up a triplet. The value of $\delta\nu_{k,1}$ is 3.77\,$\mu$Hz and The value of $\delta\nu_{k,2}$ is 6.28\,$\mu$Hz, according to Eq.\,(2). In Table 6, the $m$ $\neq$ 0 components of some modes are calculated according to the values of $\delta\nu_{k,1}$ and $\delta\nu_{k,2}$, and used to fit the other 6 observed modes in Table 1. The 25 observed modes in Table 1 are fitted in Table 6 with $\sigma_{RMS}$ = 2.59\,s. The value of $\delta\nu_{k,1}$ = 3.77\,$\mu$Hz corresponds to a rotational period of 1.54 days for R808. Bischoff-Kim (2009) identified 1 $l$ = 1 multiplet and 5 $l$ = 2 multiplets. In Table 2 of Bischoff-Kim (2009), they derived $\delta\nu_{k,1}$ = 3.59\,$\mu$Hz and $\delta\nu_{k,2}$ = 7.55\,$\mu$Hz, 8.87(26.60/3)\,$\mu$Hz, 9.62\,$\mu$Hz, 8.28\,$\mu$Hz, and 9.84\,$\mu$Hz. Those values of $\delta\nu_{k,2}$ are larger than $\delta\nu_{k,1}$/0.6 = 5.98\,$\mu$Hz.

\section{Discussion and conclusions}

The DAV star R808 was observed for 170.28 hours in April 2008 on the WET run XCOV26 (Thompson et al. 2009). There were 25 independent pulsation modes identified. We assume 19 $m$ = 0 modes and use them to constrain the fitting models. We add the core compositions of white dwarf models evolved by \texttt{MESA} to \texttt{WDEC} to obtain the thermal nuclear burning core. For the composition transition zones (C/He and He/H), we do not adopt the diffusion equilibrium profiles but element diffusion results. One group of DAV star models are evolved with the element diffusion scheme of pure Coulomb potential (Thoul, Bahcall \& Loeb 1994). The other group of DAV star models are evolved with the element diffusion scheme of screened Coulomb potential (Chen 2018, 2020). The pulsating periods of the two groups of DAV star models are calculated and used to fit the 19 observed modes.

A best fitting model is obtained with log($M_{\rm He}/M_{\rm *}$) = -2.4, log($M_{\rm H}/M_{\rm *}$) = -5.2, $T_{\rm eff}$ = 11100\,K, $M_{\rm *}$ = 0.710\,$M_{\odot}$, and log$g$ = 8.194. From the pure Coulomb potential scenario to the screened Coulomb potential scenario, the root-mean-square residual $\sigma_{RMS}$ is from 3.17\,s to 2.86\,s, 9.78\% improved. The effective temperature of our best fitting model is basically consistent with that of previous work on R808, as shown in Table 5. In Table 5, we can see that the other parameters of our best fitting model are basically consistent with that of asteroseismological work of Romero et al. (2012). The value of $\sigma_{RMS}$ for the present work is smaller than that of asteroseismological work of Castanheira \& Kepler (2009), Bischoff-Kim (2009) and Romero et al. (2012).

The observed modes of $f_{14}$, $f_{13}$, and $f_{12}$ are possible to be a triplet with $\delta\nu_{k,1}$ = 3.77\,$\mu$Hz. Therefore, the value of $\delta\nu_{k,2}$ is assumed to be 6.28\,$\mu$Hz, according to Eq.\,(2). The $m$ $\neq$ 0 components of some modes for our best fitting model are calculated according to the values of $\delta\nu_{k,1}$ and $\delta\nu_{k,2}$, and used to fit the other 6 observed modes in Table 1. Fitting the 25 observed modes, the value of $\sigma_{RMS}$ is 2.59\,s for our best fitting model. The average period spacing is 46.299\,s for $l$ = 1 modes and 25.647\,s for $l$ = 2 modes. The modes of $f_{16}$ and/or $f_{15}$ have the minimum period spacings and may be the trapped modes.

\section{Acknowledgements}

The work is supported by the foundations of NSFC of China (Grant No. 11803004 for the asteroseismological study of the screened Coulomb potential on white dwarfs and Grant No. 11563001 for the study of white dwarf pulsations). We are very grateful to X. H. Chen, Q. S. Zhang, and J. Su for their constructive suggestions.

\section*{Data availability}

The data underlying this article are available in the article and in its online supplementary material.

\label{lastpage}

\end{document}